\documentclass[%
aps,
prl,
floatfix,
amsmath,
amssymb,
superscriptaddress,
preprint
]{revtex4-2}

\usepackage{graphicx}
\usepackage{dcolumn}
\usepackage{bm}
\usepackage{hyperref}
\usepackage[mathlines]{lineno}
\usepackage[version=3]{mhchem}
\usepackage{xr}
\usepackage{float}
\usepackage[nameinlink]{cleveref}
\crefname{figure}{Fig.}{Fig.}
\Crefname{figure}{Fig.}{Fig.}
\crefname{equation}{Eq.}{Eq.}
\Crefname{equation}{Eq.}{Eq.}
\crefname{section}{Section}{Section}
\Crefname{section}{Section}{Section}
\usepackage{kotex}
\usepackage{soul}
\usepackage[dvipsnames]{xcolor}
\usepackage{siunitx}
\setstcolor{red}
\setul{}{1.5pt}

\usepackage[normalem]{ulem}

\newcommand{\Shell}{solvation shell}
\newcommand{\regionOne}{effectively ordered region}
\newcommand{\absRegionOne}{EOR}
\newcommand{\absRegionTwo}{LCR}
\newcommand{\regionTwo}{loosely correlated region}
\newcommand{\dipole}{$\theta_\text{dipole}$}
\newcommand{\criterionDistance}{$d_\textrm{LCR/EOR}$}

\newcommand{\sm}{Supplemental Material \cite{refSupplementary}}
\newcommand{\al}{\ce{Al^{3+}}}

\makeatletter
\newcommand*{\addFileDependency}[1]{
\typeout{(#1)}

\@addtofilelist{#1}
\IfFileExists{#1}{}{\typeout{No file #1.}}
}\makeatother

\newcommand*{\myexternaldocument}[1]{%
\externaldocument{#1}%
\addFileDependency{#1.tex}%
\addFileDependency{#1.aux}%
}
\myexternaldocument{supp}

\begin{document}


\title{Anomalous Water Penetration in \al~Dissolution}

\author{\firstname{Minwoo} Kim}
\author{\firstname{Seungtae} Kim}
\affiliation{School of Chemical and Biological Engineering, Institute of Chemical Processes, Seoul National University, Seoul 08826, Republic of Korea}
\author{Changbong Hyeon}
\affiliation{Korea Institute for Advanced Study, Seoul 02455, Korea}
\author{\firstname{Ji~Woong}~Yu}
\email{jiwoongs1492@kias.re.kr}
\affiliation{Korea Institute for Advanced Study, Seoul 02455, Korea}
\author{\firstname{Siyoung} Q. Choi}
\email{sqchoi@kaist.ac.kr}
\affiliation{Department of Chemical and Biomolecular engineering and KINC, KAIST, Daejeon 305-701, Korea.}
\author{\firstname{Won Bo} Lee}
\email{wblee@snu.ac.kr}
\affiliation{School of Chemical and Biological Engineering, Institute of Chemical Processes, Seoul National University, Seoul 08826, Republic of Korea}
\affiliation{School of Transdisciplinary Innovations, Seoul National University, Seoul 08826, Republic of Korea}
\date{\today}

\begin{abstract}
    The physicochemical characterization of trivalent ions is limited due to a lack of accurate force fields. By leveraging the latest machine learning force field to model aqueous \ce{AlCl3}, we discover that upon dissolution of \al, water molecules beyond the second hydration shell involve in the hydration process. A combination of scissoring of coordinating water is followed by synchronized secondary motion of water in the second \Shell~due to hydrogen bonding. Consequently, the water beyond the second solvation penetrates through the second \Shell~and coordinates to the \al. Our study reveals a novel microscopic understanding of solvation dynamics for trivalent ion.
\end{abstract}

\maketitle

Dissolution of ion in aqueous phase is a complex physico-chemical process, accompanied with the competition between ion-water and water-water dynamics. Around multivalent ion, the water dynamics and exchange kinetics are known to highly sensitive to valency and size of ions \cite{lee2017JACS}. The residence time in the first \Shell~ranges from a few $\si{\pico\s}$ to $\si{\sec}$ depending on ion species. While the stability of \al~hydrates with different coordination numbers and the energetic landscape of transitions between them have been studied, the kinetic pathway has received less attention due to unusually slow water exchange in the first \Shell~of the \al~ \cite{ohtaki1993structure,bylaska2007structure,martinez1999first,hofer2006influence,faro2010lennard,lee2017JACS}. In this letter, we investigate the kinetic pathway of hydration during the formation of hexahydrate in \ce{AlCl_3} solutions.

The complexity of interactions in aqueous solutions often requires consideration of dynamic polarization or many-body effects~\cite{wilkins2015nuclear, habershon2014zero, halgren2001polarizable, truhlar2011single, zhuang2022hydration}. Classical force fields, lacking such interactions, are not well suited to accurately predict water dynamics, even to study simple monovalent electrolyte solutions~\cite{kim2012self, avula2023understanding}. Furthermore, overestimated first \Shell~size and polarization screening significantly contribute to inaccurate descriptions of hydration shells~\cite{bedrov2019molecular, borodin2001ab, smith1997polymer, pollard2017structure, avula2023understanding}.

Despite the known limitations and the scarcity of non-polarizable models for trivalent ions, we evaluated the $12{-}6{-}4$ Lennard-Jones potential~\cite{li2015parameterization, fuhrman2023electro} for ion \Shell~structure and dynamics. This potential has shown reasonable alignment with experimental data, including hydration free energies, ion-oxygen distances, and coordination numbers for trivalent ions. However, discrepancies emerge in the second hydration shell when compared to experimental data~\cite{bol1977expShell,caminiti1979expShell,caminiti1980expShell}, which is crucial for our subsequent discussion on ion solvation.

More critically, the strong trivalent charge induces an initial overpopulation of the first \Shell~\cite{ruiz1997aluminum}, indicating that the solvation follows an associative mechanism. This observation contradicts previous reports suggesting \al~solvation exhibits a dissociative mechanism~\cite{hanauer2007searching}. Consequently, our study demands a more rigorous model. 
Extended details can be found in \sm~Section III.

\textit{Ab initio} molecular dynamics (AIMD)~\cite{iftimie2005ab}, employing quantum mechanical methods, has provided comprehensive insights into hydrolysis and revealed the structural stability of \al~hydroxy-aquo complexes ($\ce{[Al(OH)_x(H_2O)_{n-x}]^{3-x}}$, where $n = 4, 5$, and $6$)~\cite{ruiz1997aluminum}. These methods have elucidated kinetic pathways and hydration state changes in first and second hydration shells~\cite{dong2018dft, dong201927al, dong2019density, guo2023ab, ikeda2003ab,lanzani2016isomerism, jin2021theoretical, pouvreau2020mechanisms}.
However, AIMD's substantial computational demands restrict accessible spatiotemporal scales, limiting studies to highly concentrated salt solutions.

Machine learning force fields (MLFFs) represent a recent compromise between classical force fields and \textit{ab initio} methods~\cite{wu2023applications, unke2021machine, botu2017machine, poltavsky2021machine}. They offer superior computational efficiency compared to \textit{ab initio} methods, exhibiting linear scaling with system size in contrast to the cubic scaling of density functional theory (DFT). These advantages enable accurate modeling of experimental concentration domains and analysis of fluctuating coordination states beyond AIMD's typical statistical capabilities~\cite{guo2023ab, liquids2010003}.

In aqueous \ce{AlCl_3} solutions, \al\ hexahydrate (\ce{[Al(H_2O)_6]^{3+}}) is the dominant species. In this letter, we track the transient kinetic pathway to reach $n=6$ coordination. Our key discovery is a non-trivial pathway involved in the change of the coordination number from $n{=}5$ to $6$ in the early stage of ion solvation. Surprisingly, the primary source of the sixth coordinated water molecule is not the adjacent second \Shell, but a water molecule beyond the second shell. Furthermore, this sixth water molecule spends only a brief period in the second \Shell. Another key finding is that this anomalous \textit{penetration} of water in hexahydrate formation is coordinated via the hydrogen bonds (HBs) with water molecules in both the first and second shells.

\textit{Preparation ---} We employed deep potential molecular dynamics (DPMD)~\cite{zhang2018deep}, a widely used MLFF model, using the DeePMD-kit software package~\cite{wang2018deepmd}. To ensure efficient and reliable training of the MLFF model for our system, we utilized the deep potential generator (DP-GEN) framework~\cite{zhang2020dp}. The training process was conducted in a stepwise manner, starting with bulk water AIMD trajectories and snapshots provided by DP-GEN. Subsequently, the model was trained on \ce{AlCl3} solution snapshots generated by DP-GEN across a range of temperatures and concentrations. The model was trained with a loss function that was a weighted sum by energy, force, and virial, ensuring accurate performance under density modulation. The \textit{ab initio} calculations were performed using DFT with the r${^2}$SCAN functional \cite{furness2020accurate}. All molecular dynamics simulations, including DPMD, were carried out using LAMMPS~\cite{thompson2022lammps} software package, while first-principles calculations were performed using CP2K~\cite{kuhne2020cp2k}. 
Further details on the training (Section I) and validation (Section II) of our DPMD model can be found in the \sm.

A 0.1 \si{m} \ce{AlCl3} solution containing 2560 \ce{H2O} and 5 \ce{AlCl3} was relaxed using a short canonical ensemble simulation ($\sim$\si{1\ps}) at $330\,\si{\kelvin}$, followed by a $5\,\si{\ps}$ equilibration run under isobaric-isothermal conditions at $330\,\si{\kelvin}$ and $1\,\si{\bar}$. After equilibration, $15\,\si{\ps}$ production runs were performed under the same isobaric-isothermal conditions. To ensure statistical reliability, all reported results were averaged over 48 independent simulations.

\textit{Identification of hydration state ---} An \al~can form complexes with one or more \al, and the presence of such complexes can be confirmed in our system. However, for the scope of this study, we only consider monomeric \al, i.e., single \al~with their coordinating water. To identify different hydrates with varying coordination numbers, we employed Voronoi tessellation to characterize the neighbors in the first \Shell, rather than using a distance metric, which can be influenced by the cutoff choice. Water sharing a Voronoi face with the \al~were considered neighboring water. To further refine the identification, we corrected for the well-known overestimation of coordination numbers by Voronoi tessellation~\cite{yoon2019electrical}. In this correction, redundant neighbors were excluded using criteria based on solid angles, as suggested by O'Keeffe \cite{o1979proposed}. The solid angle subtended by each face generated by Voronoi tessellation was normalized by the maximum solid angle among them. Finally, water with a normalized solid angle larger than $0.5$ were retained and considered coordinating water. Note that we did not exclude hydroxide ion (\ce{OH-}) present in the system as a result of hydrolysis during hydration.

 \begin{figure}[!htp]
    \centering
    \includegraphics[width=3.4in]{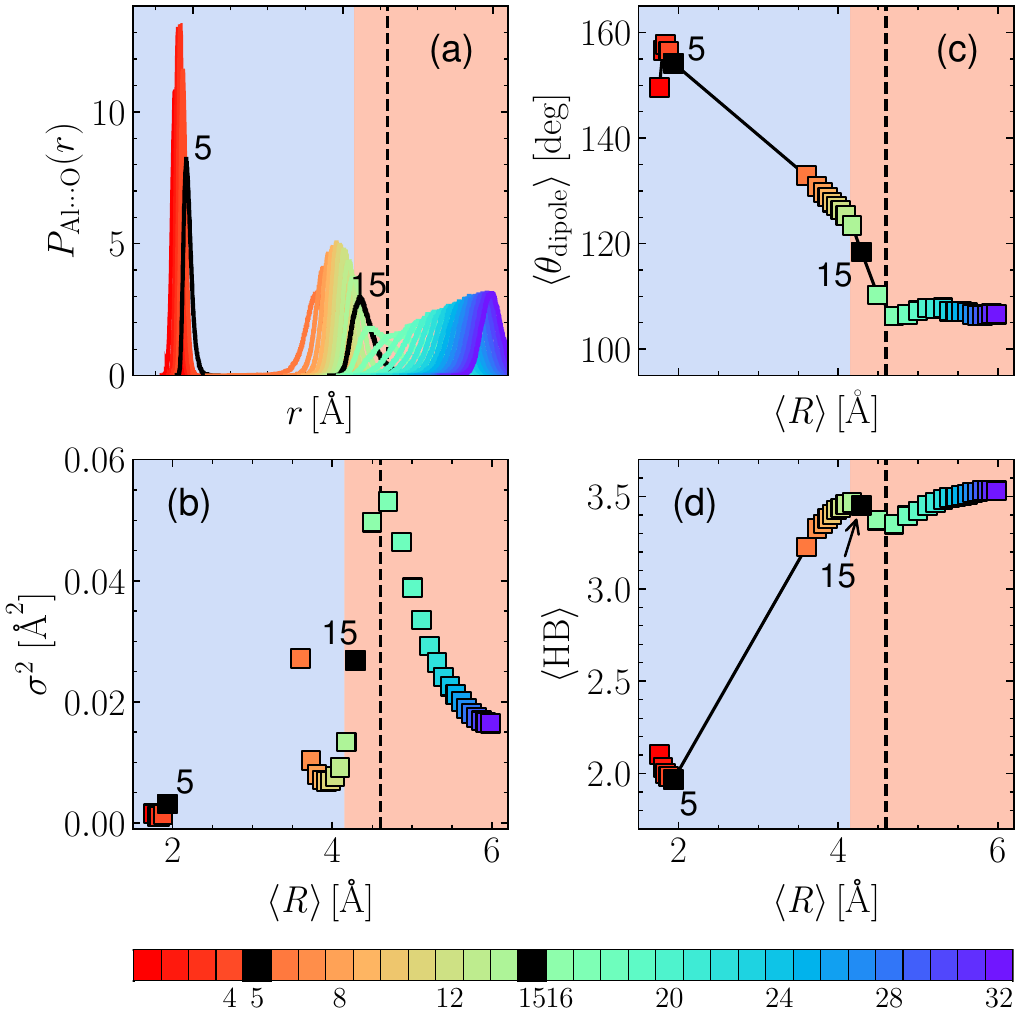}
    \caption{(a) iPDF with distance between \al\ and the \textit{i}-th neighboring oxygen, $r$. (b) iPDF variance, $\sigma^2$, (c) dipole angle, $\langle \theta_\text{dipole} \rangle$, and (d) HB number, $\langle \text{HB} \rangle$, plotted against the average distance between the \al\ and its indexed nearest oxygen neighbor, $\langle R \rangle$ (details are in \sm\ Section IV). The dashed line denotes the outer boundary of the second \Shell~at $4.6,\si{\AA}$. Background colors indicate two regions separated at a distance, \criterionDistance, of $4.15,\si{\AA}$, which denotes the boundary between \absRegionOne\ and \absRegionTwo: blue for \regionOne\ (\absRegionOne)\ and red for \regionTwo\ (\absRegionTwo). $\langle...\rangle$ denotes the statistical average over independent realizations, and over all waters or ions, and time if applicable, unless otherwise specified.}  
    \label{fig:1.Structure}
\end{figure}

\textit{Static structure ---}
We first investigate the hydration structure prior to the formation of hexahydrate. In \cref{fig:1.Structure}, the geometric structure of neighboring water when \al\ exists as a pentahydrate ($n=5$) is illustrated. The incremental probability density function (iPDF) was employed to characterize the hydration structure \cite{sharma2018born,sharma2018nature,zhuang2022hydration,avula2023understanding}. \cref{fig:1.Structure}(a) illustrates the distance iPDF between the \al\ and its 32 nearest oxygen atoms. The distribution displays two well-separated, distinct regimes: the first and the second \Shell s. The large overlap of the first five iPDF peaks around $r=1.9\,\si{\AA}$ indicates the first \Shell. After a large empty interval, the second \Shell~appears around $r=4.0\,\si{\AA}$. Unlike monovalent cations, there seems to be no overlap of iPDF between molecules in the first \Shell~and those in the second \Shell~\cite{avula2023understanding}. This can be further evident in \cref{fig:1.Structure}(b), which displays the corresponding distance variance. At the $5$th and $6$th neighbors, the variance reaches its maximum, further characterizing the solvation found in \cref{fig:1.Structure}(a). Furthermore, the low variance in the first \Shell~implies a strong coordination between water and the \al, leading to extremely slow exchange behavior of the coordinating water~\cite{lee2017JACS}. 
The clear separation of \Shell s is not limited to pentahydrates but is found in different coordination numbers as well (see \sm~Section V)


\cref{fig:1.Structure}(c) illustrates the dipole angle, $\theta_\textrm{dipole}$, as a function of the average distance between the \al\ and the indexed nearest oxygen neighbor. As a measure of orderliness, \dipole~characterizes each of the following regions. In \regionOne~(\absRegionOne), the dipole is strongly aligned towards the \al. After the \absRegionOne, shaded blue in \cref{fig:1.Structure}, \dipole\ abruptly drops and levels off at the second \Shell~limit. However, the correlation persists up to over a hundred adjacent water molecules ($\sim10\,\si{\AA}$) until it finally decorrelates (90\si{\degree}) from the \al\ (see \sm~Fig. S6). This forms another region we call \regionTwo~(\absRegionTwo), distinct from the bulk water.

\begin{figure}[!htp]
    \centering
    \includegraphics[width=3.4in]{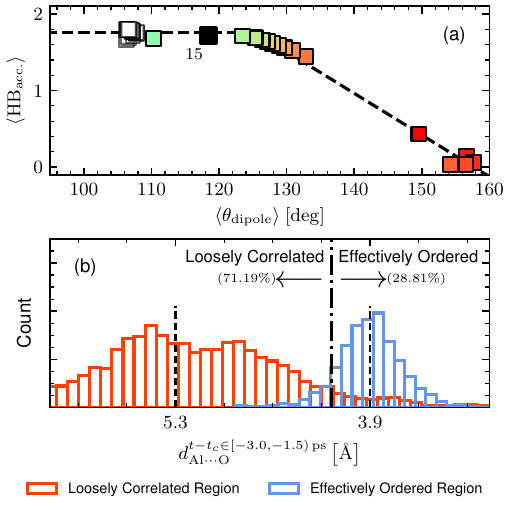}
    \caption{ (a) Correlation between $\langle \theta_\text{dipole} \rangle$ and the HB acceptor count, $\langle \text{HB}_\text{acc.} \rangle$, for the enumerated neighboring water (see \cref{fig:1.Structure} for color code). (b) Histogram of $d_{\ce{Al\bond{...}O}}$, the distance between the oxygen atom of the newly coordinating water and \al. The source region of each water is determined by comparing the time-averaged $d_{\ce{Al\bond{...}O}}$ from $-3.0$ to $-1.5\,\si{\ps}$ before coordination with \criterionDistance\ (dash-dot line). Numerical values show the average distance and population of water in each source region.}
    \label{fig:2.Corrleation}
\end{figure}

It is worth distinguishing between the second \Shell~limit and the boundary between \absRegionOne~and \absRegionTwo. The second \Shell~is determined by the variance maximum at the $n$-th neighbor and is represented as a vertical dashed line in \cref{fig:1.Structure}. In contrast, \absRegionOne~denotes the area strongly affected by the \al, with its boundary located within the second \Shell. Several different characterizations clarify distinct behaviors that mark the end of \absRegionOne. \cref{fig:1.Structure}(d) shows that the HB number reaches its maximum at the boundary between \absRegionOne~and \absRegionTwo, coinciding with \dipole~drop. Moreover, this boundary aligns with the saturation point of the HB acceptor count, as depicted in \cref{fig:2.Corrleation}(a). Beyond \absRegionOne, the HB acceptor count recovers to that of bulk water. However, \dipole~remains correlated with the \al\ even in this region.
        
\begin{figure}[!htp]
    \centering
    \includegraphics[width=3.4in]{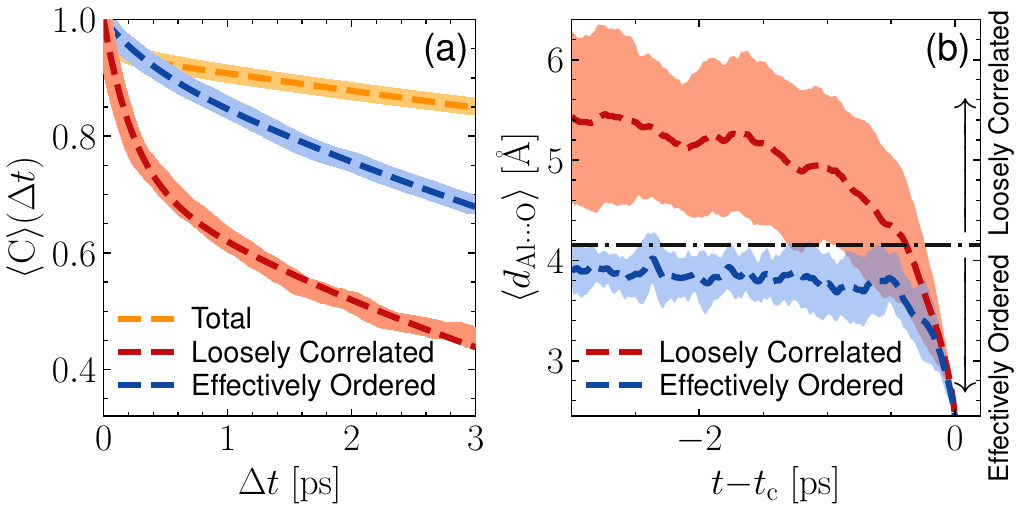}
    \caption{(a) Residence time of water in the second \Shell~of an \al. The dashed lines represent the double-exponential decay fitting, $\langle C \rangle = m \exp{(-\Delta t/\tau_1)} + (1-m) \exp{(-\Delta t/\tau_2)}$. The parameters, ($m$, $\tau_1$, $\tau_2$), are (0.939, 29.825, 0.100) for Total, (0.729, 5.898, 0.241) for LCR, and (0.937, 9.318, 0.397) for EOR. (b) $d_{\ce{Al\bond{...}O}}$ of the newly coordinating water to coordinate with it over time. The identification of source region in the legend follows the same criteria given in \cref{fig:2.Corrleation}.}
    \label{fig:3.Probability}
\end{figure}

\textit{Hydration ---} The hydration process involves the addition of a coordinating water that was not previously part of the first \Shell. To gain insight into this process, we systematically tracked the history of newly coordinating water, investigating their identity before the coordination. \cref{fig:2.Corrleation}(b) shows the population distribution of the source regions for these newly coordinated water. Notably, the major source of newly coordinated water is not \absRegionOne~but rather \absRegionTwo. This finding is unexpected, considering that water adjacent to the \al\ is more ordered and spatially closer, which should increase their likelihood of coordination.
    
To better understand the timescale of this anomalous dynamics, we estimated the residence time in the second \Shell, (\cref{fig:3.Probability}(a)). The average survival ratio $\left<C\right>{=}\left<C\right>{\left(\Delta t\right)}$ measures the fraction of water remaining in the second \Shell~after a time window, $\Delta t$. We measured $\left<C\right>$ for both all water residing in the second \Shell~and penetrating water, which exists in \absRegionTwo~among newly coordinating water. Compared to water in the second \Shell, penetrating water escapes from the second \Shell~much faster. This result rules out the possibility that water simply reside in the second coordination shell long enough to make a sequential transition to the first coordination shell. Instead, the process is abrupt. To further estimate the timescale of the process, we separated water originating from \absRegionTwo~and from those residing in \absRegionOne. \cref{fig:3.Probability}(b) shows the radial distribution of these two categories of water as a function of time. Both categories of water take less than $1\,\si{\ps}$ to coordinate after the coordination starts, i.e., when $d_\textrm{\ce{Al\bond{...}O}}$ begins to decrease. However, waters from the \absRegionTwo~approach slightly faster than those from \absRegionOne~once they cross the boundary between \absRegionOne~and \absRegionTwo. Upon observing far-away water abruptly coordinating with the \al, we term this phenomenon \textit{penetrating (hydration)}. In the rest of the letter, we will discuss the origin of this penetrating process. 
 
Recent work by Rui Shi et al.~\cite{shi2023impact} suggests that ion-specific static and dynamic properties of adjacent water around an ion result from a competition between ion-water and inter-water HBs. \cref{fig:1.Structure}(d) shows that water in the "shallow" \absRegionTwo, located near the boundary between \absRegionOne~and \absRegionTwo, exhibit a small depletion in HB count before recovering to the bulk water with $3.5$ HBs. It is worth reminding that waters in this region have an average \dipole~of ${\sim}107\si{\degree}$ against randomly oriented dipoles (${\sim}90\si{\degree}$), as shown in \cref{fig:1.Structure}(c), indicating that the ion-water correlation persists. Water in the "shallow" \absRegionTwo~lacks inter-water HBs, allowing them to move more freely under the consistent pulling by the \al\ \cite{shi2023impact}. Consequently, it is not entirely surprising that relatively distant water can coordinate the \al\ instead of water in \absRegionOne. The remaining contradiction to address is that, although water molecules in the EOR, like those in the LCR, can participate in coordination with \al\ and are expected to be the primary source of new coordination due to their stronger influence by \al, they are not actually the main source of this coordination.

\begin{figure}[!htp]
    \centering
    \includegraphics[width=3.4in]{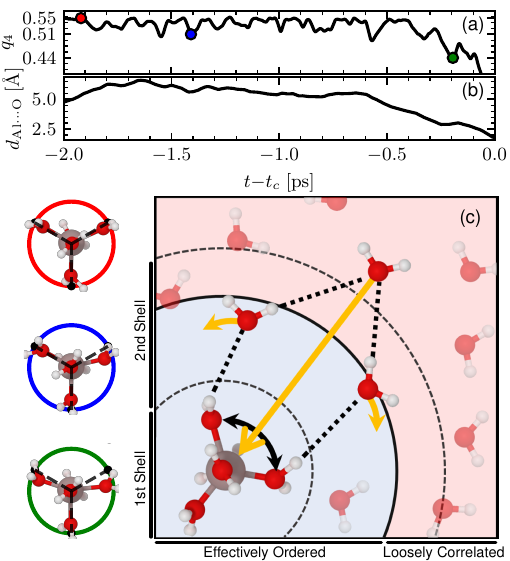}
    \caption{(a) $q_4$ for the \al hydration structure and (b) $2.0\,\si{\ps}$ history of $d_{\ce{Al\bond{...}O}}$ before coordination. Red, blue, and green dots denote the complete triangular bipyramid, the opening of two equatorial water molecules, and the fully opened state, respectively. (c) Schematic illustration of the \textit{penetrating (hydration)} process for dominant case. Other case illustrated in \sm~Fig. S16. Background colors corresponds to those in \cref{fig:1.Structure}, and dotted lines depict HBs between waters.}
    \label{fig:4.Fluctuation}
\end{figure}

To elucidate the underlying mechanism of the anomalous hydration population distribution discussed previously, we focus on the dynamics of water adjacent to the \al. \cref{fig:4.Fluctuation}(a) illustrates the orientational tetrahedral order parameter, $q_4$, over time for the \al first hydration state that is yet to be coordinated with a sixth water molecule. $q_4=0.55$ corresponds to the trigonal bipyramidal geometry of $5$ coordinating water, in contrast to $q_4=0.0$, which corresponds to the octahedral geometry of $6$ coordinating water (\sm~FIg. S8). $q_4$ fluctuates due to the scissoring motion between water molecules in the equatorial plane of the trigonal bipyramidal geometry. When $q_4\sim0.55$, the configuration of the three water molecules in the equatorial plane forms an equilateral triangle geometry (red dot in \cref{fig:4.Fluctuation}(a)). However, during the scissoring motion, the triangle geometry becomes non-regular, leading to $q_4 < 0.55$ (green dot in \cref{fig:4.Fluctuation}(a)). The widening of the water-\ce{Al}-water angle creates a vacant site to which a water molecule can coordinate. While the second coordination water can enter the site, it demands the breakage of inter-water HBs, suppressing immediate coordination.

Interestingly, the scissoring motion of two water molecules in the first \Shell~induces synchronized motion of two molecules in the second \Shell~through inter-water HBs. This synchronized motion in the second \Shell~also widens the water-\ce{Al}-water angle, as in the first \Shell, further creating room for a water molecule to coordinate. Furthermore, the resulting motion contributes to the insertion of another synchronized water molecule from the shallow \absRegionTwo, which can be confirmed by the abrupt decrease in $d_{\ce{Al\bond{...}O}}$ (\cref{fig:4.Fluctuation}(b)) following the widening of the water-\ce{Al}-water angle in the first \Shell, quantified by $q_4$ (\cref{fig:4.Fluctuation}(a)). This is the result of inter-water HBs between synchronized water in the second \Shell~and the penetrating water. Therefore, the cooperative and synchronized deshielding mechanism through a chain of inter-water HBs and under the influence of ion induced ordering explains the anomalous penetrating of distant water and abrupt coordination, as summarized in the schematic in \cref{fig:4.Fluctuation}(c).

\textit{Conclusion ---} In this work, we investigated the solvation of a trivalent ion, which is challenging to study, through DPMD simulations. By employing a large system that allows for the study of \al at a scale practically inaccessible to AIMD, we report an unconventional hydration dynamics of of \al\ by distant water. We reveal that this anomalous penetration dynamics of water results from a complex cooperative motion of water molecules in the first to the exterior of second shell, facilitated by a chain of inter-water HBs and by ion induced electrostatic correlation. The insights gained from this study would be of great help in understanding ion solvation in various contexts, such as catalysis, ion transport, and material design~\cite{meng2022low, zhang2016maleic, berben2015catalysis}. Finally, the methodological advancements presented in this work open up new possibilities for investigating complex ion solvation phenomena that were previously challenging to explore.

\begin{acknowledgments}
    This work was supported by Samsung Science and Technology Foundation (project no. SRFC-MA2201-03) and by the Ministry of Education of the Republic of Korea and the National Research Foundation of Korea (RS-2024-00343871). This work was supported by a KIAS Individual Grant (AP091501 to J.W.Y., CG035003 to C.H.) at Korea Institute for Advanced Study. We especially thank Tae Jun Yoon for his fruitful discussion. 
    
    Minwoo Kim and Seungtae Kim contributed equally to this work. 
\end{acknowledgments}
\bibliography{ref}
\nocite{hendrycks2016gaussian}
\nocite{soper2008quantum}
\nocite{Lemmon2024nist}
\nocite{zhang2021phase}
\nocite{bol1977expShell}
\nocite{caminiti1979expShell}
\nocite{caminiti1980expShell}
\nocite{ohtaki1993expShell}
\nocite{HUMP96}
\end{document}


\title{Supplementary Material for "Anomalous Water Penetration in \ce{Al^{3+}} Dissolution"}

\author{Minwoo Kim}
\affiliation{School of Chemical and Biological Engineering, Institute of Chemical Processes, Seoul National University, Seoul 08826, Republic of Korea}

\author{Seungtae Kim}
\affiliation{School of Chemical and Biological Engineering, Institute of Chemical Processes, Seoul National University, Seoul 08826, Republic of Korea}

\author{Changbong Hyeon}
\affiliation{Korea Institute for Advanced Study, Seoul 02455, Korea}

\author{Ji~Woong~Yu}
\email{jiwoongs1492@kias.re.kr}
\affiliation{Korea Institute for Advanced Study, Seoul 02455, Korea}

\author{Siyoung Q Choi}
\email{sqchoi@kaist.ac.kr}
\affiliation{Department of Chemical and Biomolecular engineering and KINC, KAIST, Daejeon 305-701, Korea.}

\author{Won Bo Lee}
\email{wblee@snu.ac.kr}
\affiliation{School of Chemical and Biological Engineering, Institute of Chemical Processes, Seoul National University, Seoul 08826, Republic of Korea}
\affiliation{School of Transdisciplinary Innovations, Seoul National University, Seoul 08826, Republic of Korea}

\date{\today}

\maketitle

\section{\label{S1.Train} Training Process of Machine Learning Force Fields (MLFFs)}

\subsection{\label{S1.A.AIMD} Ab initio molecular dynamics (AIMD) simulations}

\begin{table}[H] 
    \centering
    \begin{tabular}{c|c|c|c|c|c}
         \hline nSystem &  Ensemble & Composition (\ce{AlCl3} : \ce{H2O}) & Temp [\si{\kelvin}] & Pressure [\si{\bar}]/Box [$\si{\AA}$] & Simulation time [\si{\ps}] \\
         \hline
         1 & NpT & 0 : 64 & 330 & 1.000 & 3 \\ 
         3 & NVT & 0 : 64 & 330 & 12.425 & 0.5 \\ 
         3 & NVT & 0 : 64 & 500 & 12.425 & 0.5 \\ 
         3 & NVT & 0 : 64 & 1000 & 12.425 & 0.5 \\ 
         1 & NVT & 0 : 64 & 2000 & 12.425 & 0.5 \\
         \hline
    \end{tabular}
    \caption{Simulation information which are generated for bulk water system.}
    \label{tab:data4water}
\end{table}

\begin{table}[H] 
    \centering
    \begin{tabular}{c|c|c|c|c|c}
         \hline nSystem &  Ensemble & Composition (\ce{AlCl3} : \ce{H2O}) & Temp [\si{\kelvin}] & Pressure [\si{\bar}]/Box [$\si{\AA}$] & Simulation time [\si{\ps}] \\
         \hline
         4 & NVT & 1 : 64 & 330 & 12.700 & 0.5 \\ 
         4 & NVT & 1 : 64 & 500 & 12.700 & 0.5 \\ 
         4 & NVT & 1 : 64 & 1000 & 12.700 & 0.5 \\ 
         4 & NVT & 1 : 64 & 2000 & 12.700 & 0.5 \\ 
         4 & NVT & 2 : 64 & 330 & 12.593 & 0.5 \\ 
         4 & NVT & 2 : 64 & 500 & 12.593 & 0.5 \\ 
         4 & NVT & 2 : 64 & 1000 & 12.593 & 0.5 \\ 
         4 & NVT & 2 : 64 & 2000 & 12.593 & 0.5 \\ 
         1 & NVT & 3 : 64 & 500 & 13.713 & 0.5 \\ 
         1 & NVT & 3 : 64 & 1000 & 13.713 & 0.5 \\ 
         1 & NpT & 1 : 64 & 330 & 1.000 & 5.0 \\ 
         1 & NpT & 2 : 64 & 330 & 1.0003 & 10.0 \\ 
         \hline
    \end{tabular}
    \caption{Simulation information which are generated for \ce{AlCl3} aqueous solution system.}
    \label{tab:data4alcl3}
\end{table}

 Table \ref{tab:data4water}, \ref{tab:data4alcl3} shows the list of initial condition of AIMD simulation, before using DP-GEN active learning like process.
 All the simulations were conducted with using CP2K programs.
 Each simulation were conducted with 0.5 \si{\fs} time step for Born-Oppenheimer molecular dynamics and 1200 Ry used for electron density cutoffs.
 A triple-$\zeta$ valence and polarization (TZV2P) basis set was used for all atoms in this simulation.
 And the core electrons were treated with Geodecker-Teter-Hutter (GTH) norm-conserving pseudopotentials.
 The revsised version of strongly constrained and appropriately normed (SCAN) functional, which known as r$^2$SCAN was used.

\subsection{\label{S1.B.DataSet} Deep Potential GENerator (DP-GEN)}
    DP-GEN was a active learning like process framework, which was used to train the machine learning force fields.
    DP-GEN process was constructed with big three step, train, exploring and labeling.
    
    In the train step, deep potential smooth edition (DeepPot-SE) with "two-atom embedding descriptor" were used to represent condensed aqueous electrolytes.
    Sel parameters were selected with "H: 96, O: 52, Al: 12, Cl:24" 
    Gaussian Error Linear Units (GELUs)\cite{hendrycks2016gaussian} were used for activation function, with a parameter as "gelu\_tf".
    The loss function were constructed with force, energy and virial.
    Each value were 0.2, 1000, 0.2.

    In the exploring step, the molecular dynamics (MD) simulation were conducted with LAMMPS.
    We set the initial condition of MD simulation with various density and temperature for exploring the broad phase space.
    Such as for bulk water system, initial configuration were random configuration with density from 0.1 to 1.2 g/cm$^3$.
    Temperature was set from $300$ to $2000\,\si{\kelvin}$.
    And for \ce{AlCl3} aqueous solution, density, temperature and composition (\ce{AlCl3}:\ce{H2O}) were set from 1.0 to 1.5 g/cm3, from 300 to 1000 \si{\kelvin} and from 1:64 to 3:64.
    Furthermore, dimeric hydrated \ce{Al^{3+}} complex was a key for training.
    So, additional data was generated with AIMD simulation.
    Initial configuration were made from DP-GEN process.
    We select standard as maximum force deviation (Eq. \ref{eqr:max}) with lower bound as 0.11 and upper bound 0.30.
    
    \begin{equation}
        \eta = \max \sqrt{|\left< \boldsymbol{F}_i - \left< \boldsymbol{F}_i \right> \right>|^2}
        \label{eqr:max}
    \end{equation}

    In DP-GEN process, 4 machine learning force fields were generated with different seed and batch size. 
    $\boldsymbol{F}_i$ is ith machine learning force fields forces of each simulation snapshot.
    $\left<\right>$ is a symbol of ensemble the machine learning force fields.

    In the labeling step, CP2K program was used for single point energy calculation.
    The simulation setting was same as AIMD simulation (Supplementary Section \ref{S1.A.AIMD}).
    Because of changed CP2K logging format, we modify the \texttt{cp2k} module of \texttt{dpdata} included in \texttt{DPGEN} package (https://github.com/deepmodeling/dpdata.git) to extract the data for training from CP2K2022.2 simulation results.

\clearpage
\section{\label{S2.Valid} Validation of the trained MLFF}
    \begin{figure}[H] 
        \centering
        \includegraphics[width=0.6\linewidth]{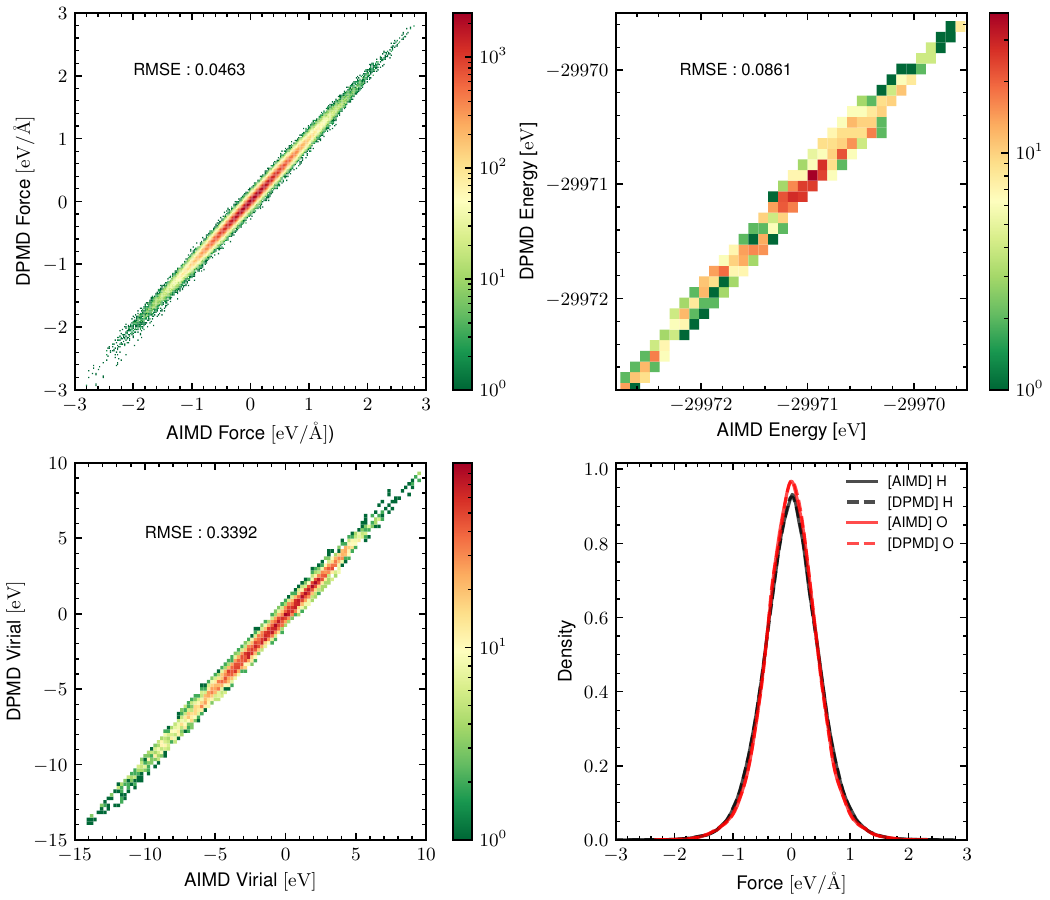}        \caption{
        Compare the force, energy, virial and atomic force distribution of AIMD and DPMD simulation results. Root-mean-square-error (RMSE) were used for calculate the error deviation. 64 \ce{H2O} were included in the system with temperature of 330 \si{\kelvin} and pressure of 1 \si{\bar} with NpT ensemble.
        }
    \end{figure}
    
    \begin{figure}[H] 
        \centering
        \includegraphics[width=0.6\linewidth]{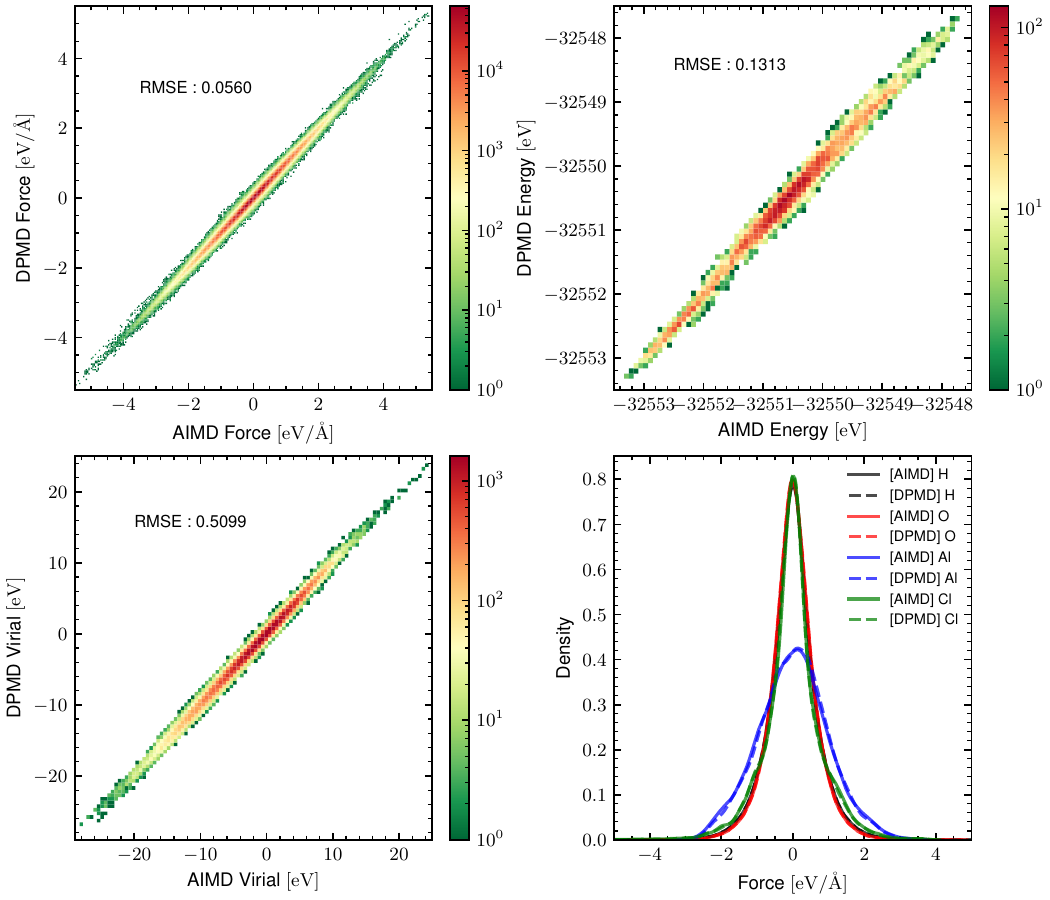}        \caption{
        Compare the force, energy, virial and atomic force distribution of AIMD and DPMD simulation results. Root-mean-square-error (RMSE) were used for calculate the error deviation. 2 \ce{AlCl3} and 64 \ce{H2O} were included in the system with temperature of 330 \si{\kelvin} with NVT ensemble.
        }
    \end{figure}
    
    \begin{figure}[H] 
        \centering
        \includegraphics[width=\linewidth]{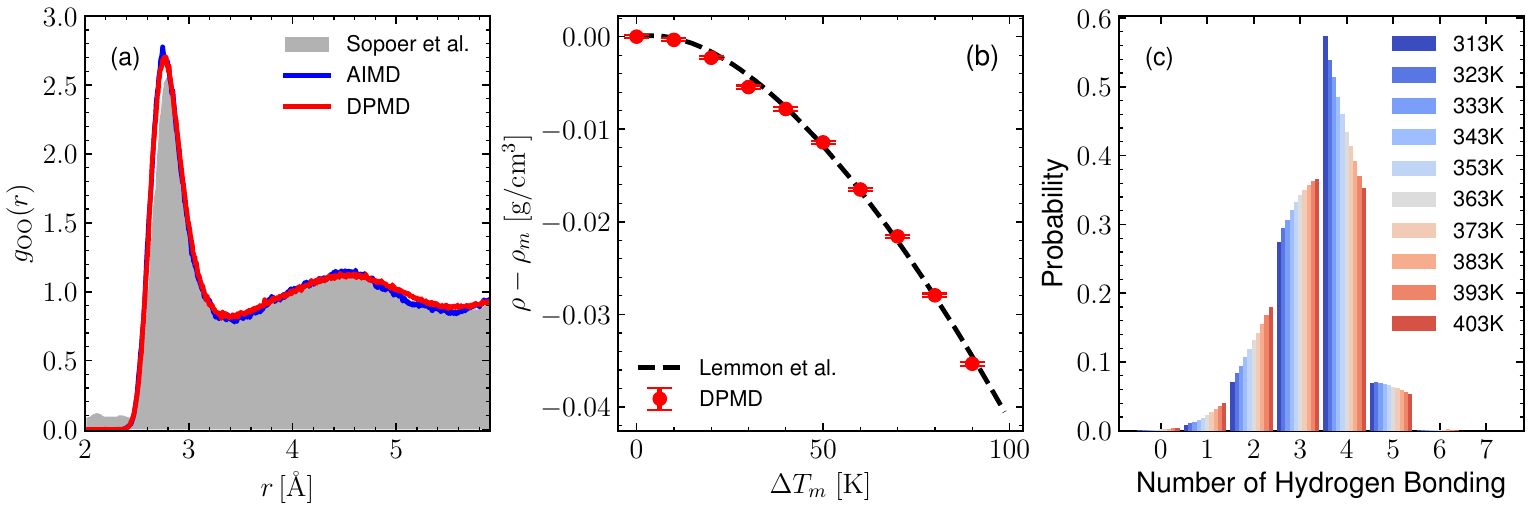}
        \caption{
        DPMD means deep potential molecular dynamics which were simulated with previous trained MLFF.
        AIMD means ab intio molecular dynamics which data were generated by CP2K program.
        (a) Oxygen-Oxygen radial distribution function were calculated from same initial configuration which conditions are temperature as 330 \si{\kelvin}, pressure as 1 \si{\bar}, ensemble as isobaric-isothermal(NpT) ensemble and 64 \ce{H2O}.
        Sopoer et al. suggested experimental data from X-ray scattering results \cite{soper2008quantum}.
        (b) Compare the density of each temperature between results of DPMD simulations and reference data \cite{Lemmon2024nist},
        DPMD simulations were conducted with 512 \ce{H2O} and 1 \si{\bar}.
        In the figure, $\rho_m$ mean density at melting temperature.
        $\Delta T_m$ means $T - T_m$, $T_m$ is melting temperature.
        Melting temperatures of each data were $273.15 \si{\kelvin}$ for reference and $313 \si{\kelvin}$ for DPMD based on SCAN xc-functional \cite{zhang2021phase}.
        (c) The number of hydrogen bonding changed with temperature.
        The data were generated with 512 \ce{H2O} and 1 \si{\bar} DPMD simulation.
        }
        \label{fig:enter-label}
    \end{figure}
    
    \begin{figure}[H] 
        \centering
        \includegraphics[width=1.0\linewidth]{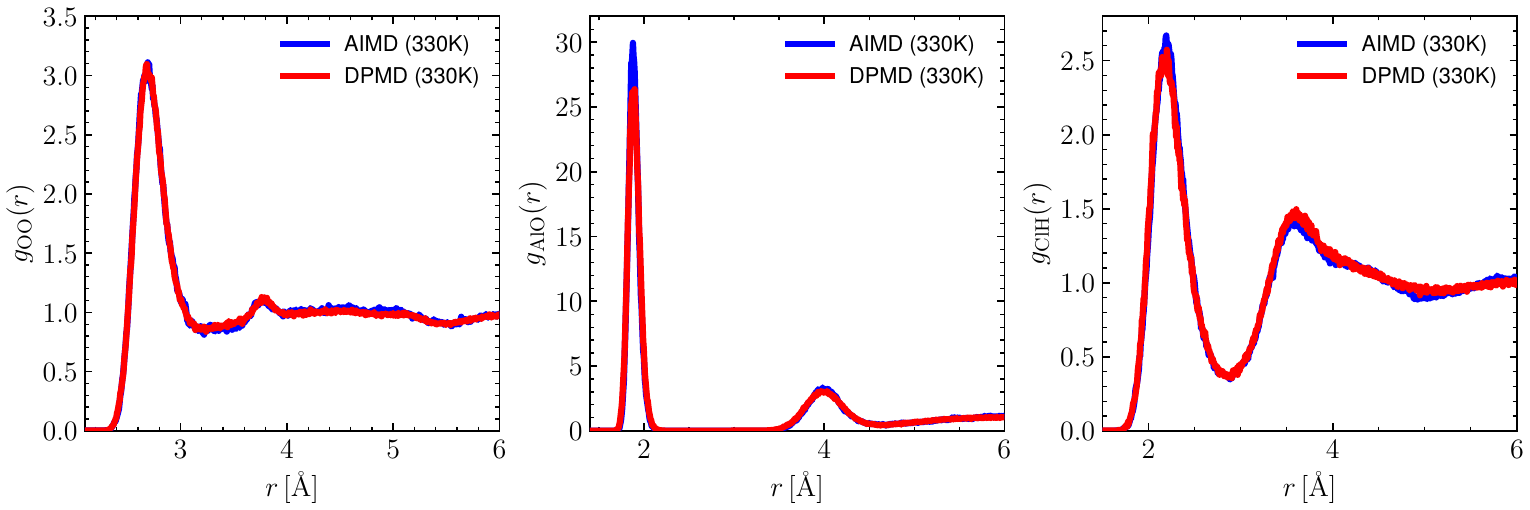}
        \caption{Compare the radial distribution function (RDF) between results of AIMD and DPMD.
        Canonical ensemble (NVT) simulation were conducted with 330 \si{\kelvin} temperature and 12.5926 \si{\AA}\ box until 20 \si{\ps} simulation time.
        The RDFs which depict (a) Oxygen-Oxygen (b) \ce{Al^{3+}}-Oxygen (c) Chlorine-Hydrogen were calculated after 10 \si{\ps} simulation time.}
        \label{fig:enter-label}
    \end{figure}
\clearpage
\section{\label{S.CMD} Comparison with Conventional Molecular Dynamics (CMD)}
    CMD refers to conventional molecular dynamics simulation using the 12-6-4 Lennard-Jones (LJ) type nonbonded model~\cite{li2015parameterization}. For the CMD simulation, a 0.1 $\si{m}$ \ce{AlCl3} salt solution, consisting of 2560 \ce{H2O} and 5 \ce{AlCl3}, was run under isobaric-isothermal conditions at 298 \si{\kelvin} and 1 \si{\bar} for 10 \si{\ns}. The production of the simulation was calculated after an initial equilibration period of 5 \si{\ns}. On the other hand, DPMD refers to deep potential molecular dynamics. For the DPMD simulation, a 0.1 \si{m} \ce{AlCl3} salt solution, also containing 2560 \ce{H2O} and 5 \ce{AlCl3}, was run under isobaric-isothermal conditions at 330 \si{\kelvin} and 1 \si{\bar} for 1 \si{\ns}. The production phase of this simulation was calculated after an initial equilibration period of 0.5 \si{\ns}.
    \begin{figure}[!htp]
        \centering
        \includegraphics[width=0.6\linewidth]{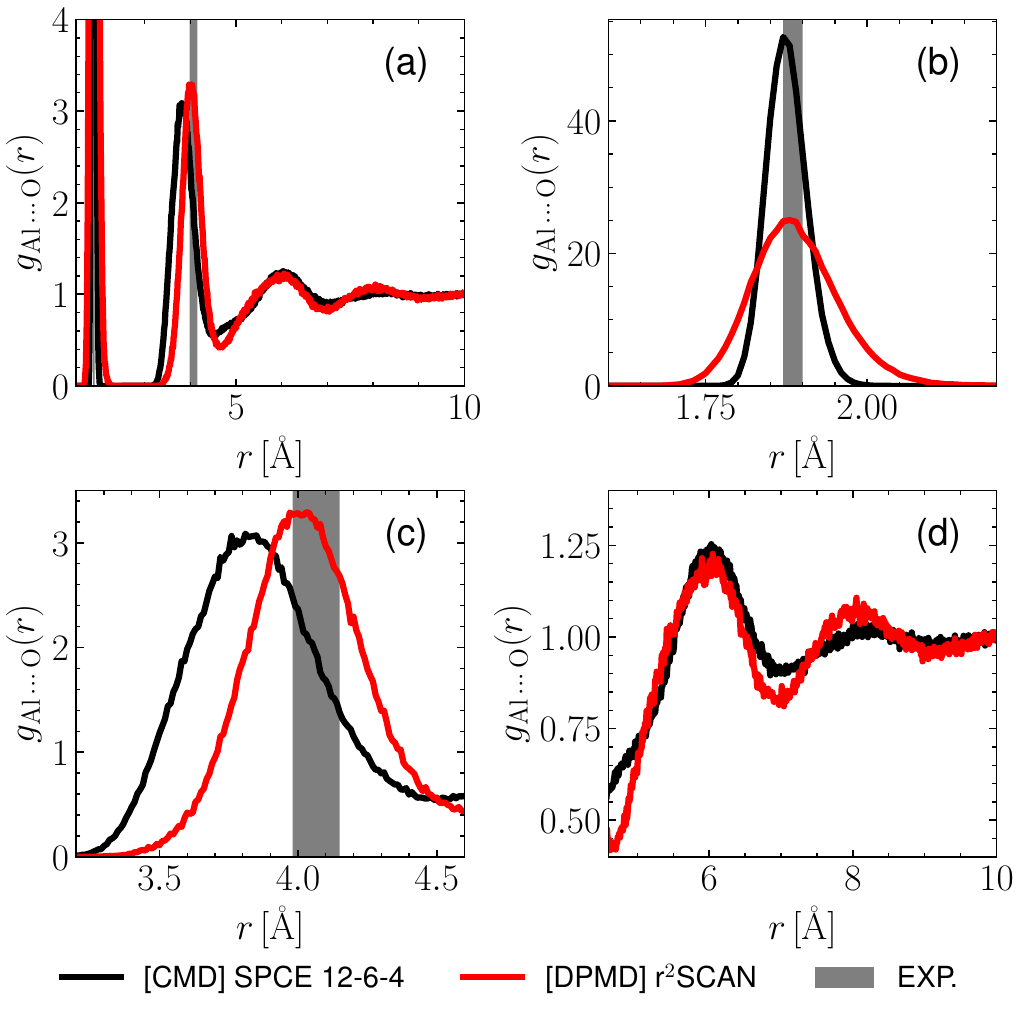}
        \caption{$g_{\ce{Al\bond{...}O}}(r)$ demonstrate that radial distribution function between \ce{Al^{3+}} and oxygen in water. EXP. denotes the experiment data. \ce{Al\bond{...}O} 1st peak (1.882, 1.902) \cite{caminiti1979expShell}, (1.90) \cite{bol1977expShell}, (1.87) \cite{caminiti1980expShell}, (1.88) \cite{ohtaki1993expShell}. \ce{Al\bond{...}O} 2nd peak (4.01, 4.02) \cite{caminiti1979expShell}, 4.10-4.15 \cite{bol1977expShell}, 3.99 \cite{caminiti1980expShell}.}
        \label{fgr:CMDvsDPMD_RDF}
        \label{fgr:SupFig_CMDvsDPMD_RDF}
    \end{figure}
    
    \begin{figure}[!htp]
        \centering
        \includegraphics[width=0.6\linewidth]{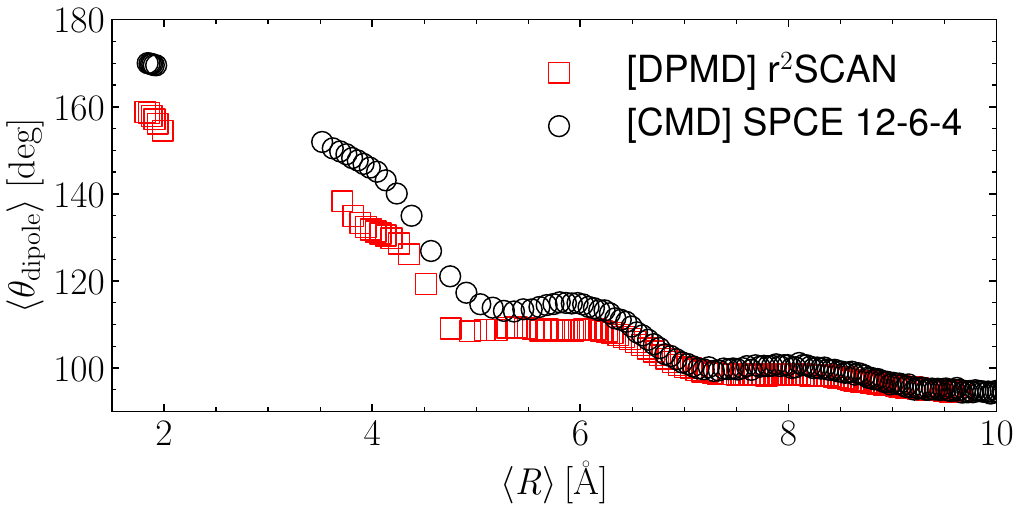}
        \caption{Average dipole angle, $\langle \theta_\text{dipole}\rangle$, were computed by \eqref{eqr:dipole3}, $\langle R \rangle$, between \ce{Al^{3+}} and oxygen atom of its 128 neighbor water molecules. Over 128, the dipole angle of water molecules are uncorrelated with \ce{Al^{3+}}.}
        \label{fgr:longdipoleangle}
    \end{figure}
    
    \begin{figure}[!htp]
        \centering
        \includegraphics[width=1.0\linewidth]{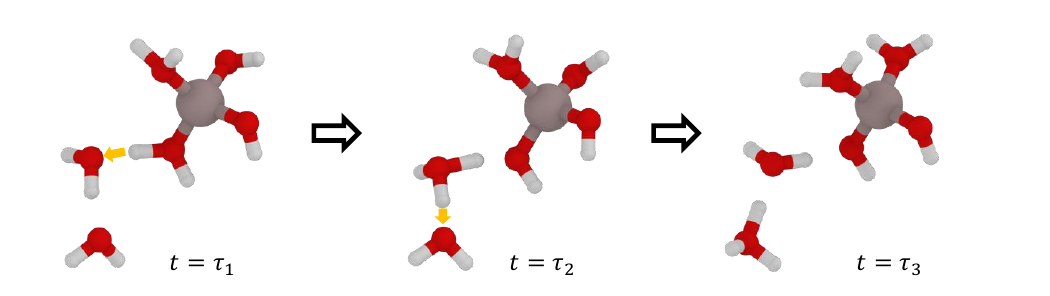}
        \caption{Snapshot over time for the hydrolysis reaction of water in 1st hydration shell of \ce{Al^{3+}}, in case of 4-coordinate.}
        \label{fig:enter-label}
    \end{figure}
    
\clearpage
\section{\label{S.equations}Computational Metrics for Structural and Orientational Analysis}
\subsection{\label{S.iPDF} Increase Probability Density Function (iPDF)}

    Increase probability density function (iPDF or P(r)) is defined by Eq. \eqref{eqr:iPDF}.
    In the equation, (n) means nth-oxygen from each \ce{Al^{3+}}.
    i means ith-\ce{Al^{3+}} and j means its nth-oxygen.
    $\delta$(r) is Dirac delta function about one dimensional case.
    $N_{\text{Al}}$ is the number of \ce{Al^{3+}}.
    $r_{ij}$ is the distance between atom i and j.
    And $\left< \right>$ means the ensemble average. 
    
    \begin{equation}
        P_{Al \cdots O}^{(n)}(r) = \frac{1}{N_{Al}} \left< \sum_{i}^{N_{Al}} \delta(r-r_{ij}) \right>
        \label{eqr:iPDF}
    \end{equation}

\subsection{\label{S.DipoleAngle} Dipole angle between \ce{Al^{3+}} and water molecules}

    \begin{equation}
        \hat{\mu}_{\text{nH} \cdots \text{O}} = \frac{\sum_i^{n} \left( \textbf{x}_{\text{H}_i} - \textbf{x}_{\text{O}} \right)}{\lVert  \sum_i^{n} \left( \textbf{x}_{\text{H}_i} - \textbf{x}_{\text{O}} \right) \rVert}
        \label{eqr:dipole1}
    \end{equation}
    
    \begin{equation}
        \hat{\mu}_{\text{Al} \cdots \text{O}} = \frac{\left( \textbf{x}_{\text{Al}} - \textbf{x}_{\text{O}} \right)}{\lVert  \left( \textbf{x}_{\text{Al}} - \textbf{x}_{\text{O}} \right) \rVert}
        \label{eqr:dipole2}
    \end{equation}
    
    \begin{equation}
        \theta_{\text{dipole}} = \arccos \left ( {\hat{\mu}_{\text{OHn}} \cdot \hat{\mu}_{\text{Al} \cdots \text{O}}} \right )
        \label{eqr:dipole3}
    \end{equation}
    
    Eq. \eqref{eqr:dipole3} was used to calculate the dipole angle between two unit dipole moment.
    One of the unit dipole moment, defined in Eq. \eqref{eqr:dipole1}, was for dipole moment of \ce{OH_n} molecules.
    $n$ in Eq. \eqref{eqr:dipole1} means that the number of hydrogen atom.
    For example, hydronium ion (\ce{H3O+)} of $\mathrm{n}$ is 3.
    The other was dipole moment, defined in Eq. \eqref{eqr:dipole2}, of \ce{Al^{3+}} and oxygen atom in water molecules.
    In the equation, $\hat{\mu}$ means that unit vector of dipole moment.
    $\lVert \textbf{a} \rVert$ is magnitude notation of vector \textbf{a}.
    The position of atom was denoted by \textbf{x}$_{i}$.
    Dot product of two vector is denoted by $\textbf{a} \cdot \textbf{b}$.

\subsection{\label{S.HBonds} The number of hydrogen bonding count}
    The hydrogen bond, $h_{ij}$, defined by under condition.
    Distance between two different oxygen atom is within 3.5 \AA.
    And the angle of HOO should be less than $30\deg$.
    So, the ith water's hydrogen bonding number is $H_i$.
    
    \begin{equation}
        h_{ij} = 
        \begin{cases}
            1 & \text{if the condition are met.}\\
            0 & \text{else.}
        \end{cases}
    \end{equation}
    
    \begin{equation}
        H_i = \sum_{j}^{N} h_{ij}
    \end{equation}
    
\clearpage
\subsection{\label{S.Orientational} Orientational tetrahedral order parameter ($q_4$)}
    Orientational tetrahedral order parameter was denoted as $q_4$, which is defined
    Eq. \eqref{eqr:q4}. In the equation, $\theta_{ijk}$ is angle between oxygen atom i, j and k-th \ce{Al^{3+}}. \textrm{cn} means that the coordination number of Oxygen atom in first hydration shell of \textit{k}-th \ce{Al^{3+}}. The order parameter is closer to 1.0 as the structure approaches the tetrahedron.
    
    \begin{equation}
        q_{4,k} = 1 - \frac{3}{8} \sum_{j=1}^3 \sum_{j+1}^{\textrm{cn}} \left( \cos \theta_{ijk} + \frac{1}{3} \right)^2
        \label{eqr:q4}
    \end{equation}
    
    \begin{figure}[H] 
        \centering
        \includegraphics[width=0.8\linewidth]{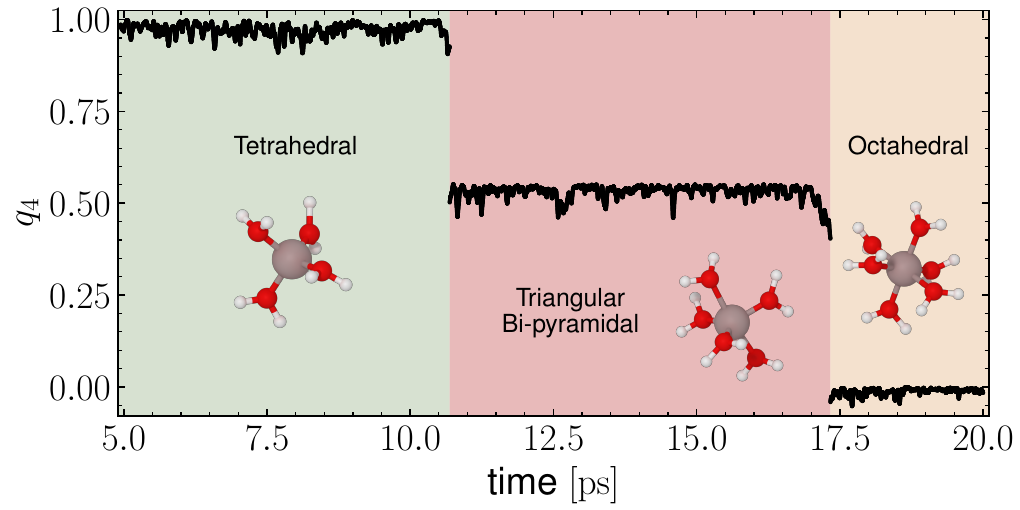}
        \caption{The above figure shows the orientational tetrahedral order parameter ($q_{4,k}$) according to the time flow of k-th \ce{Al^{3+}}. Eq. \eqref{eqr:q4} was used to calculate these order parameters. The three areas separated by color represent three distinct hydration states. In the case of tetrahedral structure, fluctuation is performed near 1.0, fluctuation is performed near 0.55 in the case of triangular bi-pyramidal structure, and finally fluctuation is performed near 0.00 in the case of octahedral structure.}
        \label{fgr:q4}
    \end{figure}

\clearpage
\section{\label{S4.Others} Other hydration state}

\subsection{4-coordinate}
    \begin{figure}[!htp]
        \centering
        \includegraphics[width=0.7\linewidth]{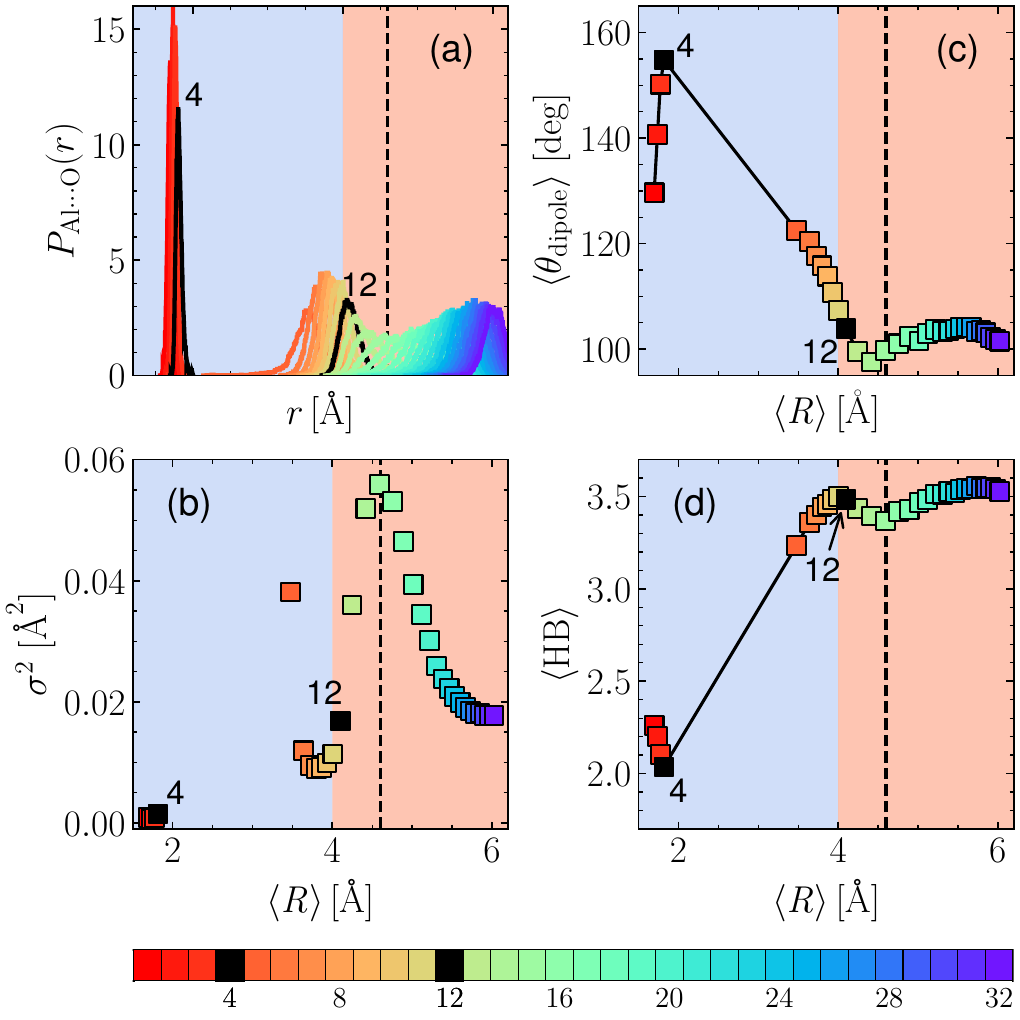}
        \caption{
            These results were computed for the 32 neighboring oxygen atoms closest to the \ce{Al^{3+}}. (a) Incremental Probability Density Function (iPDF), (b) iPDF variance, (c) average dipole angle, and (d) average number of HBs (equations are in Supplemental Material Section \ref{S.equations}) are plotted against the average distance between the \ce{Al^{3+}} and its \textit{i}-th neighboring oxygen. The dashed line denotes the outer boundary of the second solvation shell at $4.6\,\si{\AA}$. The background colors within each figure serve as visual aids for two distinct region separated by a distance of $4.15,\si{\AA}$, with blue indicating effectively ordered regions and red denoting intermediate regions. These figures represent results for the 4-coordinate case.}    
        \label{fgr:CN4_Fig1}
    \end{figure}
    
\clearpage
\subsection{6-coordinate}
    \begin{figure}[!htp]
        \centering
        \includegraphics[width=0.7\linewidth]{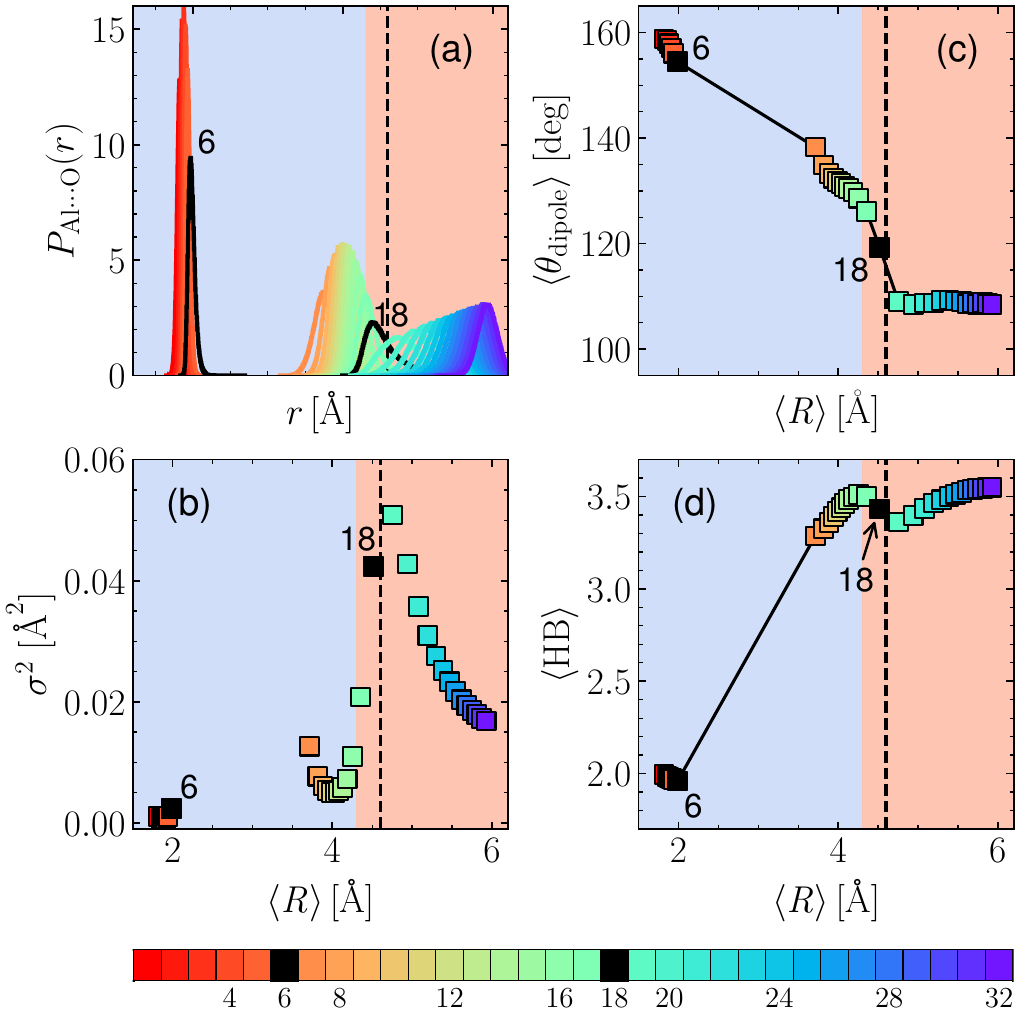}
        \caption{
            These results were computed for the 32 neighboring oxygen atoms closest to the \ce{Al^{3+}}. (a) Incremental Probability Density Function (iPDF), (b) iPDF variance, (c) average dipole angle, and (d) average number of HBs (equations are in Supplemental Material Section \ref{S.equations}) are plotted against the average distance between the \ce{Al^{3+}} and its \textit{i}-th neighboring oxygen. The dashed line denotes the outer boundary of the second solvation shell at $4.6\,\si{\AA}$. The background colors within each figure serve as visual aids for two distinct region separated by a distance of $4.15,\si{\AA}$, with blue indicating effectively ordered regions and red denoting intermediate regions. These figures represent results for the 6-coordinate case.}
        \label{fgr:CN6_Fig1}
    \end{figure}
\clearpage
    \begin{figure}[!htp]
        \centering
        \includegraphics[width=\linewidth]{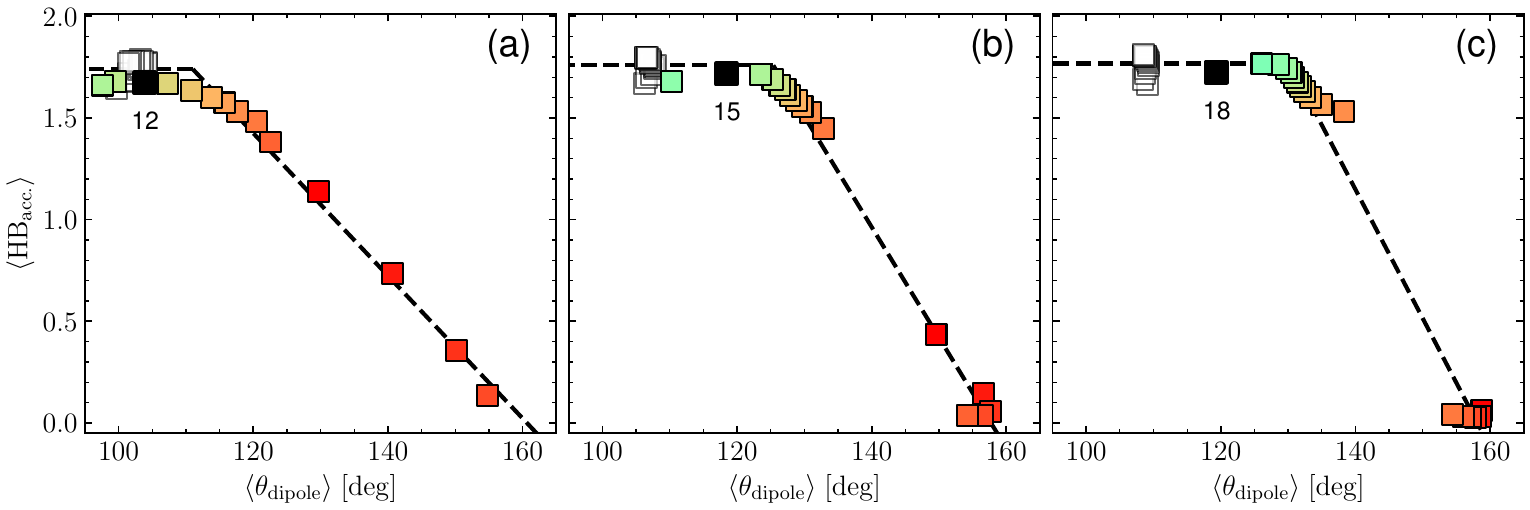}
        \caption{
        The dipole angle and hydrogen bonding number represent the average values computed from the 32 nearest oxygen atoms of water surrounding the \ce{Al^{3+}}. For the hydrogen bonding number, only cases meeting the acceptance criteria are considered. The guideline presented corresponds to the result of the average hydrogen bond number line of water located outside the second hydration shell, along with the linear regression of water within the linear section.
        Within the figure, colored squares denote water molecules situated inside the second hydration shell, whereas other squares represent those located outside the second hydration shell. Each figure shows cases of each coordination number, (a) 4 (b) 5 (c) 6}
        \label{fig:enter-label}
    \end{figure}
    
\clearpage
\section{Anomalous Penetration from 4-coordinate to 5-coordinate}
    \begin{figure}[!htp]
        \centering
        \includegraphics[width=0.5\linewidth]{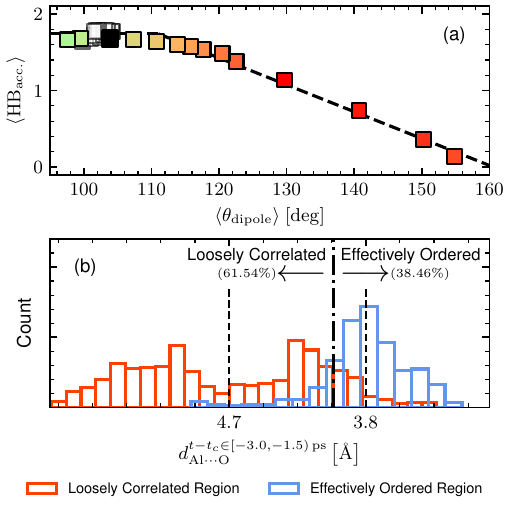}
        \caption{(a) The correlation between the average dipole angle and the average HB acceptor count for the ith neighboring water molecule. Colored points denote water molecules situated inside the second hydration shell, whereas other squares represent those located outside the second hydration shell. The scatter plot uses the same color scheme as \cref{fgr:CN4_Fig1}, with dark points representing the 12th neighbor. The dashed lines indicate the limit of HB acceptor counts and the results of linear regression for the effectively ordered region's scatter points. (b) The histogram of the distance between the newly coordinating water molecule and \ce{Al^{3+}}. The region, in which the newly coordinating water belong, was determined based on the average distance between $1.5\,\si{\ps}$ and $3.0\,\si{\ps}$ before it coordinates with \ce{Al^{3+}}. Here, "newly coordinating water" refers to the water molecule that will coordinate with the \ce{Al^{3+}}. The dash-dot line at $4.00\,\si{\AA}$ is the criterion for separating the two regions. Approximately, $61.54$\% of the newly coordinating water molecules, transitioning from 4 to 5-coordinate states, existed in the intermediate region. Additionally, the average distances in each area are $4.7\,\si{\AA}$ and $3.8\,\si{\AA}$, respectively.}
        \label{fgr:subfig_2}
    \end{figure}
    \begin{figure}[!htp]
        \centering
        \includegraphics[width=0.5\linewidth]{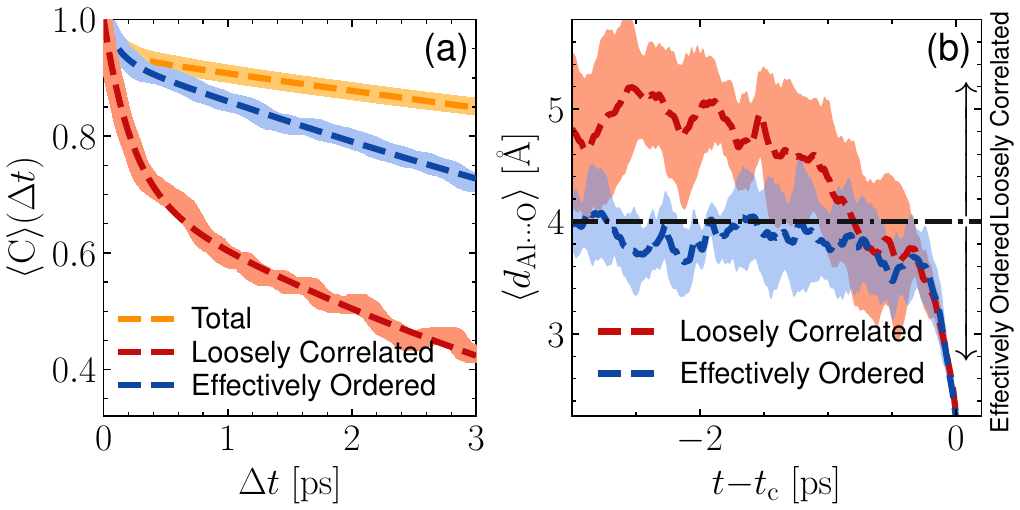}
        \caption{In the figures Loosely Correlated and Effectively Ordered means that the region of newly coordinating water, which coordination number change from 4 to 5. (a) Residence time of water in the second \Shell~of \ce{Al^{3+}}. The dashed lines represent the double-exponential decay fitting, $\langle C \rangle = m \exp{(-\Delta t/\tau_1)} + (1-m) \exp{(-\Delta t/\tau_2)}$. The parameters, ($m$, $\tau_1$, $\tau_2$), are (0.939, 29.825, 0.100) for Total, (0.715, 5.743, 0.223) for LCR, and (0.935, 11.955, 0.119) for EOR. (b) $d_{\ce{Al\bond{...}O}}$ of the newly coordinating water to coordinate with it over time. The identification of source region follows the same criteria given in \cref{fgr:subfig_2}.}
        \label{fgr:subfig_3}
    \end{figure}

\clearpage
\section{\label{S.AlOdistance} Species Distribution and Distance}    
\begin{figure}[!htp]
        \centering
            \includegraphics[width=0.6\linewidth]{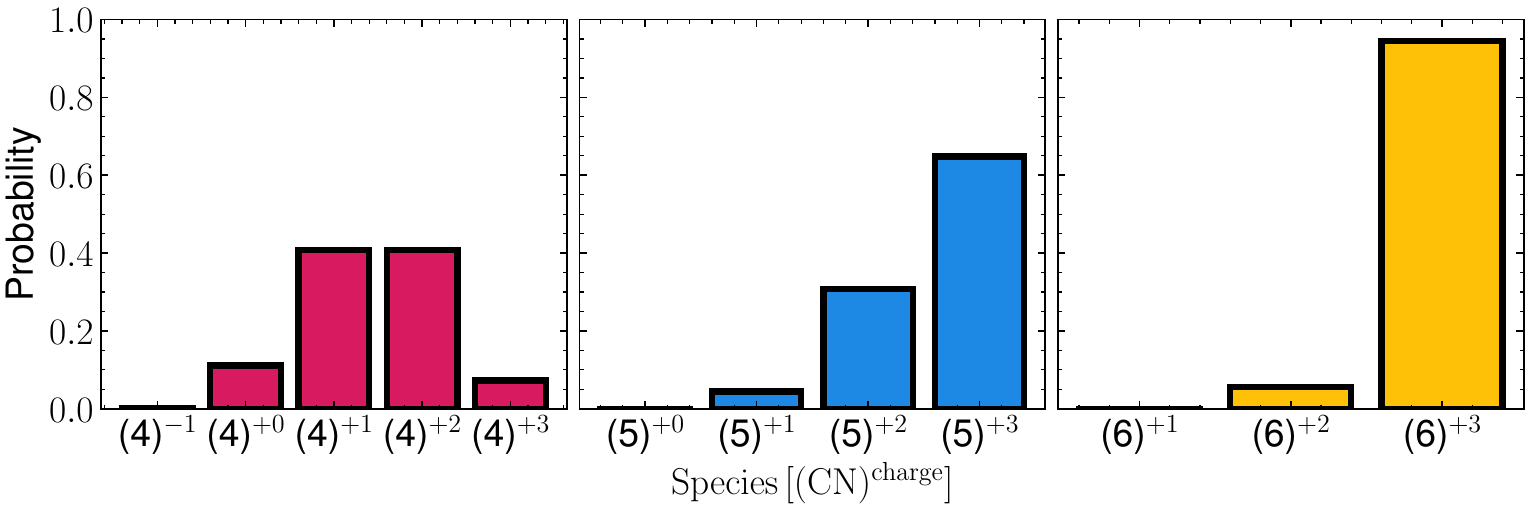}
        \caption{Species($(n)^{x}$) probability of the 48 ensemble simulation. The $(n)^{x}$ denotes $n$-th coordinate with $x$ charge state.}
        \label{fgr:speciesDistribution}
    \end{figure}

\begin{figure}[!htp]
        \centering
            \includegraphics[width=0.6\linewidth]{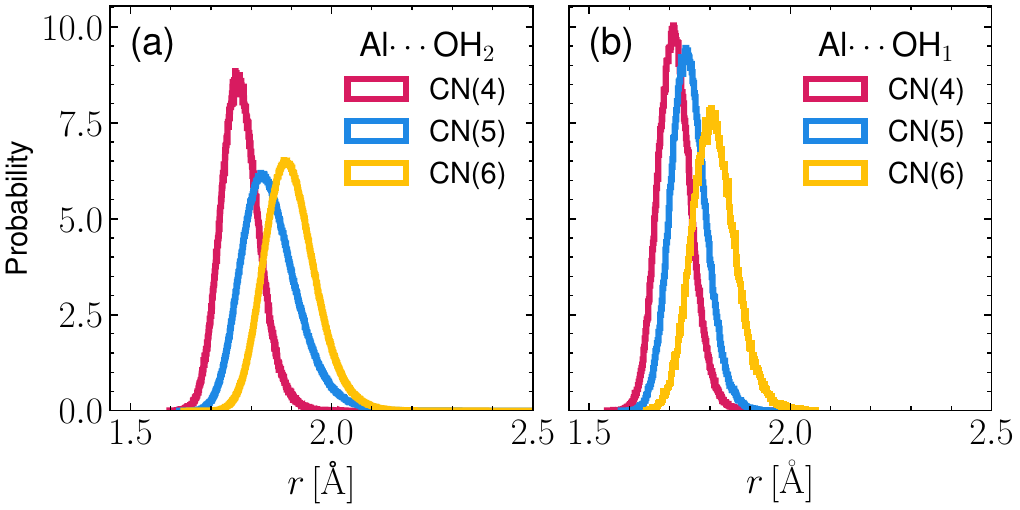}
        \caption{This figure shows that the distance density function between \ce{Al^{3+}} and oxygen atom in its 1st hydration shell. Of the above two figures, (a) is the case of \ce{H2O}, and (b) is the case of \ce{OH-}.}
        \label{fgr:AlOdistance}
    \end{figure}

    \cref{fgr:speciesDistribution} shows that hydrolysis occur during our simulation. And the probability distributions are depends on the coordination number. And \cref{fgr:AlOdistance} shows that the distance distribution between \ce{Al^{3+}} and its first shell coordinating oxygen atom are also depend on the coordination number.

\clearpage
\section{\label{S.AlOdistance} Penetration Process Schematics}
\begin{figure}[H]
    \centering
    \includegraphics{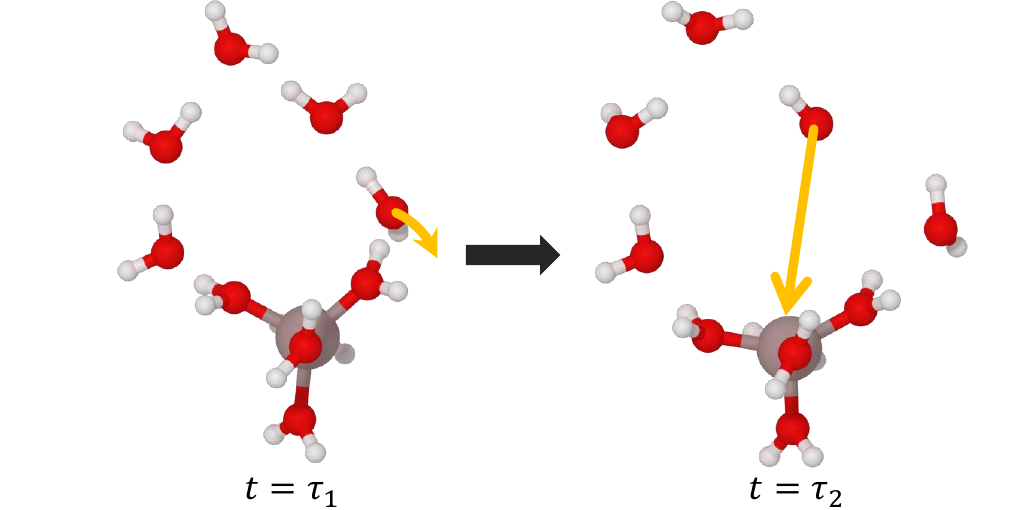}
    \caption{Schematic snapshots of the \textit{penetration} process in case of 7 \ce{H2O} correlated. Dotted-line means hydrogen bonding. we only visualized the important and water nearby \ce{Al^{3+}}.}
    \label{fgr:SupFig_HBn7}
\end{figure}

\section{\label{S.MOVIE} Supplementary Videos}
In the supplementary videos, we use Visual Molecular Dynamics (VMD) \cite{HUMP96} to visually illustrate the important interactions between the \ce{Al^{3+}} and the surrounding water molecules. The visualization focuses on key water molecules near the \ce{Al^{3+}}. Specifically, the central green sphere represents the \ce{Al^{3+}}. Water molecules are shown with yellow-colored oxygen atoms, which serve as gatekeepers of the \ce{Al^{3+}} hydrated complex. Red springs indicate dynamic hydrogen bonding interactions between the water molecules, while water molecules with blue-colored oxygen atoms represent those newly coordinating with the \ce{Al^{3+}}.

\bibliography{ref}